\documentclass[aps,showpacs,twocolumn,floats,superscriptaddress,prl,Letter]{revtex4}
\usepackage{graphicx}
\usepackage{amsmath}
\usepackage{amssymb}
\usepackage{mathtext}

\begin{document}

\newcommand {\e} {\varepsilon}
\newcommand {\ph} {\varphi}
\newcommand {\Nt} {{\mbox{\scriptsize N}}}
\newcommand {\Ct} {{\mbox{\scriptsize C}}}
\newcommand {\rmP} {\mathrm{P}}
\renewcommand {\Re} {\mathrm{Re}}
\renewcommand {\Im} {\mathrm{Im}}
\renewcommand {\d} {\mathrm{d}}

\title{Reverse Engineering of Proteasomal Translocation Rates}

\author{D.S.\ Goldobin}
\affiliation{Department of Physics, University of Potsdam,
        Postfach 601553, D--14415 Potsdam, Germany}
\affiliation{Department of Theoretical Physics, Perm State University,
        15 Bukireva str., 614990, Perm, Russia}
\author{M.\ Mishto}
\affiliation{CIG, University of Bologna,
        I--40126 Bologna, Italy}
\affiliation{Institute of Biochemistry, Charit\'e, Humboldt University, Monbijoustr.\ 2, 10117 Berlin, Germany}
\author{K.\ Textoris-Taube}
\author{P.M.\ Kloetzel}
\affiliation{Institute of Biochemistry, Charit\'e, Humboldt University, Monbijoustr.\ 2, 10117 Berlin, Germany}
\author{A.\ Zaikin}
\affiliation{Departments of Mathematics and IFWH,
             University College London, Gower street, WC1E~6BT London, UK}

\begin{abstract}
We address the problem of proteasomal protein translocation and
introduce a new stochastic model of the proteasomal digestion
(cleavage) of proteins. In this model we account for the protein
translocation and the positioning of cleavage sites of a
proteasome from first principles. We show by test examples and by
processing experimental data that our model allows reconstruction
of the translocation and cleavage rates from mass spectroscopy
data on digestion patterns and can be used to investigate the
properties of transport in different experimental set-ups.
Detailed investigation with this model will enable theoretical
quantitative prediction of the proteasomal activity.
\end{abstract}

\maketitle

A macromolecular complex, the proteasome, is the central molecular
machine for the degradation of intracellular
proteins~\cite{1994_Rock}. Proteasomes have a pivotal role in
antigen processing that prepares epitopes for an immune
system~\cite{2001_Kloetzel_nature}. They exist in cells as the
free proteolytically active core, the barrel-shaped 20S proteasome
(Fig.\,\ref{fig1}b), and as associations of this core with
regulatory complexes (PA700 or PA28) at its
ends~\cite{2000_Tanahashi}. Here we consider {\it in vitro}
proteasomal digestion assays widely used in molecular biology and
immunology to investigate proteasomes.

\begin{figure}[!t]
\center{
  \includegraphics[width=0.3995\textwidth]%
 {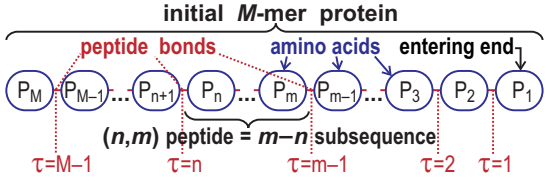}
}
  \caption{Peptide bonds are indexed with $\tau$, aa with $\rmP_{\!i}$}
  \label{fig1a}
\end{figure}

\begin{figure}[!b]
\center{
  \includegraphics[width=0.3995\textwidth]%
 {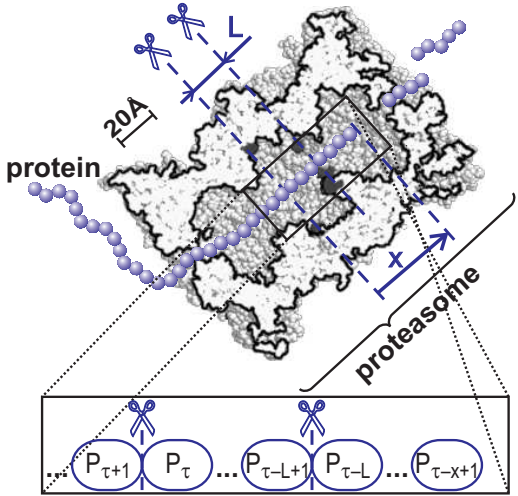}
}
  \caption{Infiltration of a protein strand into the 20S proteasome: The
scissors mark the positions of cleavage sites rings at $x=0$ and
$x=L$; the cleavage occurs via the attaching-detaching of the
protein to cleavage sites (dark-grey color); thus, in the figure
the bonds between $\rmP_{\!\tau+1}$ and $\rmP_{\!\tau}$ and
between $\rmP_{\!\tau-L+1}$ and $\rmP_{\!\tau-L}$ may be cleaved. The first aa has  index $\tau-x+1$ after $\tau-x$
aa have  been cut out (see inset).}
  \label{fig1}
\end{figure}

A protein (Fig.\,\ref{fig1a}) enters the proteasome and is translocated into the
central chamber where it is cleaved into fragments by  the
cleavage sites. We assume that 6 cleavage sites are arranged along two rings (Fig.\,\ref{fig1}).
Fragments of the protein produced are removed through proteasome
gates. The translocation proteasomal function can qualitatively
change the expression of the specific fragment, {\it e.g.}, an
epitope, because modified translocation and thus increased time of
residence near the cleavage terminal changes the conditions of
cleavage. Moreover, impairment of proteasomal degradation,
probably due to translocation malfunction, might contribute to the
pathology of various neurodegenerative
conditions~\cite{2006_Rubinsztein_nature}.

The mechanism of protein translocation remains unknown. It is also
unknown whether translocation properties are different for
different proteasome types (constitutive or immuno-), with/without
different regulatory complexes, and with different experiment
conditions (concentration ratios, temperature, {\it etc.}). Only a
few papers address the translocation problem but these are either
based on semi-phenomenological descriptions of uptaking and
translocation of the
protein~\cite{2000_Holzhuetter_bj,2002_Peters,2005_Luciani} or
they suggest a transport mechanism hypothesis not yet verified
experimentaly~\cite{2005_Zaikin_proteasome}. On the other hand,
there exist several
facilities~\cite{2000_Kuttler_Paproc,2005_Tenzer} to predict where
the protein will be cleaved but as numerous experiments
show~\cite{2008_Mishto_Zaikin_proteamalg} these algorithms do not always
work reliably. The reason is that these algorithms utilize
experimental data resulted eventually from some specific protein
sequence and translocation function, but the prediction is made
based only on the sequence, ignoring the translocation function.
 In contrast to these
approaches, here we introduce a stochastic model which allows one
to {\it reconstruct} both the translocation and cleavage rates
from mass spectroscopy (MS) data on digestion patterns.
Collecting the reconstructed features of a specific proteasome
type can be used for a reliable prediction of the fragment
expression.

In our model of protein translocation and degradation by the
proteasome  we assume that:

\noindent
(1)\;The event rate of the protein shift by one amino acid (aa)
into the proteasome (to the right in Fig.\,\ref{fig1}) depends
only on the length $x$ of the protein forward end beyond the
cleavage sites nearest to the proteasome chamber entrance used
for protein infiltration (the left ones in Fig.\,\ref{fig1}); this
event rate is given by the translocation rate function (TRF)
$v(x)\equiv v_x$. The backward motions of the entering strand are
neglected. These assumptions do not impose significant
restrictions on the physical mechanism of the translocation
process: they are valid for the Brownian drift in a tilted
spatially-periodic potential~\cite{2001_Reimann} as well as for
the ratchet effect~\cite{2005_Zaikin_proteasome}, {\it etc}. The
TRFs of different proteasome species (20S, 26S,
$\pm$PA28~\cite{2000_Tanahashi}) may differ.

\noindent
(2)\;When the protein strand is close to the cleavage site, the
event rate of the cleavage depends on the sequence of aa nearest
to the peptide bond cleaved~\cite{2005_Tenzer}. For the given
protein, this conditional cleavage rate (CCR),
$\gamma(\tau)\equiv\gamma_\tau$, is a function of the bond number
$\tau$ (Fig.\,\ref{fig1a}); later on we use $\tau$ near the first
ring of cleavage sites as a {\em time-like variable}.

\noindent
(3)\;The peptides (cleaved parts of the protein degraded) leave
the chamber through proteasome gates. Due to their
mobility being higher in comparison to that of the protein, processed
peptides leave the chamber quick enough to neglect both their
possible further splitting and their influence on the protein transport.

Let us now introduce the distribution $w(x|\tau)$ which is the
probability of the protein forward end beyond the first ring of
the cleavage sites to be of the length $x$, when the $\tau$th
bond is near that ring, in our terms, at the discrete ``time
moment'' $\tau$. We measure $x$ in aa, thus $x$ and $\tau$ are
integer. To describe the ``temporal'' evolution of distribution
$w(x|\tau)$, we consider the shift of the protein strand into the
proteasome for one aa, {\it i.e.}, the transition $\tau\to\tau+1$.
Let us decompose $w(x|\tau+1)$ as
\[\textstyle
w(x|\tau+1)=\sum_j w_j(x|\tau+1)\,,
\]
where $w_j(x|\tau+1)$ are the contributions due to different
scenarios of this transition. Along with $w(x|\tau)$, we account
$Q(n,m|\tau)$, the amount of the peptide $(n,m)$, which is the
$m$--$n$ subsequence of the degraded protein (Fig.\,\ref{fig1}),
generated during transition $\tau\to\tau+1$.

There are three possible elementary events:
\\
(a)\;the protein strand shift: $x\to x+1$, $\tau\to\tau+1$; the
event rate is $v_x$;
\\
(b)\;the cleavage on the first ring of cleavage sites ($x=0$):
$x\to 0$, $\tau\to\tau$; the event rate is $\gamma_\tau$;
\\
(c)\;the cleavage on the second ring of cleavage sites ($x=L$,
$L$ is the distance between the rings of cleavage sites,
Fig.\,\ref{fig1}): $x\to L$, $\tau\to\tau$; the event rate is
$\gamma_{\tau-L}$.

In terms of these elementary events the possible scenarios of
transition $\tau\to\tau+1$ are

\noindent
1)\;{\em Elementary event~(a)}. Its probability is
\[
P_1(x|\tau)=v_x/\big(v_x+\gamma_\tau+\Theta(x\!-\!L\!-\!1)\gamma_{\tau-L}\big),
\]
where the Heaviside function $\Theta(x\!<\!0)=0$,
$\Theta(x\!\ge\!0)=1$. In this scenario, $x\to x+1$, and
\begin{equation}
w_1(x+1|\tau+1)=P_1(x|\tau)\,w(x|\tau)\,.
\label{eq_aux-1w}
\end{equation}
No peptides are generated;

\noindent
2)\;{\em Elementary event~(b)}, which may not be followed by
anything but the strand shift by one aa (as there is nothing to be
cleaved). This scenario probability is
\[
P_2(x|\tau)=\gamma_\tau/\big(v_x+\gamma_\tau+\Theta(x\!-\!L\!-\!1)\gamma_{\tau-L}\big).
\]
In this scenario, $x\to 1$, and
\begin{equation}
\textstyle
w_2(x|\tau+1)=\delta_{x,1}\sum_{x'=1}^\infty
P_2(x'|\tau)\,w(x'|\tau)\,.
\label{eq_aux-2w}
\end{equation}
The peptides cut out are
\begin{equation}
Q_2(\tau,\tau-x+1|\tau)=P_2(x|\tau)\,w(x|\tau)\,;
\label{eq_aux-2Q}
\end{equation}

\noindent
3)\;{\em Elementary event~(c), which may be followed either by
strand shift~(1) or by scenario~(2)}. The probability of the first
stage (c) is
\[
P_c(x|\tau)=\Theta(x\!-\!L\!-\!1)\,\gamma_{\tau-L}/\big(v_x+\gamma_\tau+\gamma_{\tau-L}\big).
\]
After event~(c), when $x\to L$, the number of the system states
generated is
\[
\textstyle
w_c(x|\tau)=\delta_{x,L}\sum_{x'=L+1}^\infty
P_c(x'|\tau)\,w(x'|\tau)\,,
\]
and the peptides cut out are
\[
Q_c(\tau-L,\tau-x+1|\tau)=P_c(x|\tau)\,w(x|\tau)\,.
\]
The subsequent events~(1) or (2) should be regarded as the
respective above mentioned scenarios starting with distribution
$w_c(x|\tau)$, {\it i.e.},
\begin{eqnarray}
 &&\hspace{-5mm}\textstyle
 w_{c1}(x|\tau+1)=P_1(L|\tau)\,w_c(x-1|\tau)\nonumber\\[3pt]
 &&\textstyle
 =P_1(L|\tau)\;\delta_{x,L+1}\sum_{x'=L+1}^\infty P_c(x'|\tau)\,w(x'|\tau)\,,
\label{eq_aux-c1w}
 \\[6pt]
 &&\hspace{-5mm}\textstyle
 Q_{c1}(\tau\!-\!L,\tau\!-\!x\!+\!1|\tau)
 =P_1(L|\tau)\,Q_c(\tau\!-\!L,\tau\!-\!x\!+\!1|\tau)\nonumber\\[3pt]
 &&\qquad\textstyle
 =P_1(L|\tau)\,P_c(x|\tau)\,w(x|\tau)\,,
\label{eq_aux-c1Q}
 \\[6pt]
 &&\hspace{-5mm}\textstyle
 w_{c2}(x|\tau+1)=\delta_{x,1}\sum_{x'=1}^\infty P_2(x'|\tau)\,w_c(x'|\tau)\nonumber\\[3pt]
 &&\textstyle
 =\delta_{x,1}\;P_2(L|\tau)\sum_{x'=L+1}^\infty P_c(x'|\tau)\,w(x'|\tau)\,,
\label{eq_aux-c2w}
 \\[6pt]
 &&\hspace{-5mm}\textstyle
 Q_{c2}(\tau\!-\!L,\tau\!-\!x\!+\!1|\tau)=P_2(L|\tau)\,Q_c(\tau\!-\!L,\tau\!-\!x\!+\!1|\tau)\nonumber\\[3pt]
 &&\qquad\textstyle
 =P_2(L|\tau)\,P_c(x|\tau)\,w(x|\tau)\,.
\label{eq_aux-c2Q-1}
 \\[6pt]
 &&\hspace{-5mm}\textstyle
 Q_{c2}(\tau,\tau-x+1|\tau)=P_2(x|\tau)\,w_c(x|\tau)\nonumber\\[3pt]
 &&\textstyle\qquad
 =\delta_{x,L}\,P_2(L|\tau)\sum_{x'=L+1}^\infty P_c(x'|\tau)\,w(x'|\tau)\,.
\label{eq_aux-c2Q-2}
\end{eqnarray}

Collecting
Eqs.\,(\ref{eq_aux-1w}),(\ref{eq_aux-2w}),(\ref{eq_aux-c1w}),(\ref{eq_aux-c2w}),
we find master equation
\begin{eqnarray}
&&w(1|\tau+1)
 =\sum\limits_{x=1}^{L}\frac{\gamma_\tau\,w(x|\tau)}{v_x+\gamma_\tau}
 +\hspace{-5pt}\sum\limits_{x=L+1}^{\infty}\hspace{-2pt}\frac{\gamma_\tau\,w(x|\tau)}{v_x+\gamma_\tau+\gamma_{\tau-L}}
\nonumber\\
&&
 \qquad+\frac{\gamma_\tau}{v_L+\gamma_\tau}
 \sum\limits_{x=L+1}^{\infty}\frac{\gamma_{\tau-L}\,w(x|\tau)}{v_x+\gamma_\tau+\gamma_{\tau-L}};\label{eq01-1}\\[5pt]
&&w(L+1|\tau+1)
 =\frac{v_L}{v_L+\gamma_\tau}\nonumber\\
&& \qquad\times\biggl\lbrack w(L|\tau)
 +\sum\limits_{x=L+1}^{\infty}\frac{\gamma_{\tau-L}\,w(x|\tau)}{v_x+\gamma_\tau+\gamma_{\tau-L}}
 \biggr\rbrack;\label{eq01-2}\\[5pt]
&&\hspace{-4mm}x\ne 1,\;x\ne L+1\,:
\nonumber\\
&&w(x|\tau+1)
 =\frac{v_{x-1}\,w(x-1|\tau)}{v_{x-1}+\gamma_\tau+\Theta(x\!-\!L\!-\!1)\gamma_{\tau-L}}.\label{eq01-3}
\end{eqnarray}
Here $x=1,2,3,...,M$ and $\tau=1,2,3,...,M-1$, where $M$ is the
length of the protein (Fig.\,\ref{fig1}).


\begin{figure}[!b]
\center{
  \includegraphics[width=0.47\textwidth]%
 {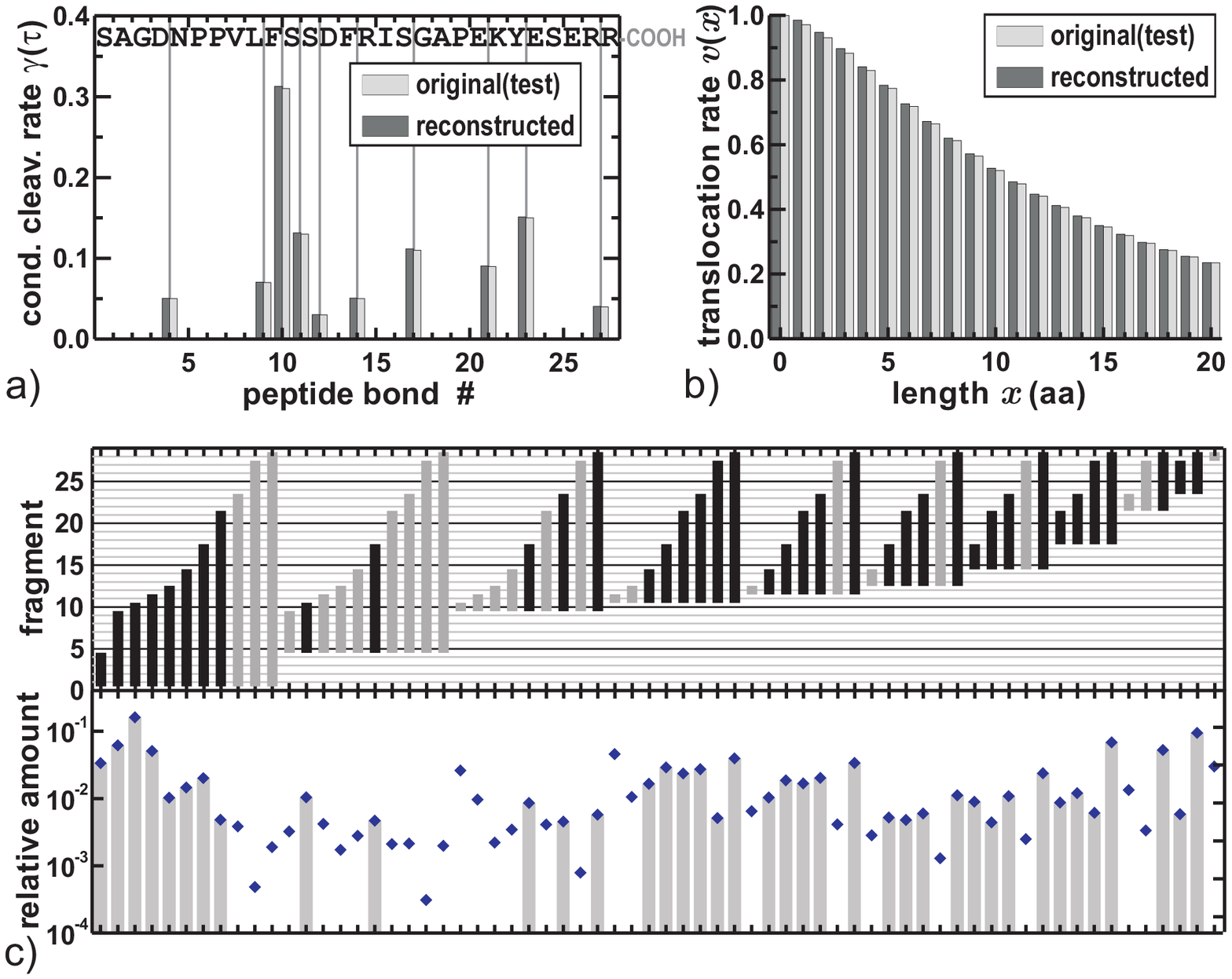}
\\[5pt]
  \includegraphics[width=0.47\textwidth]%
 {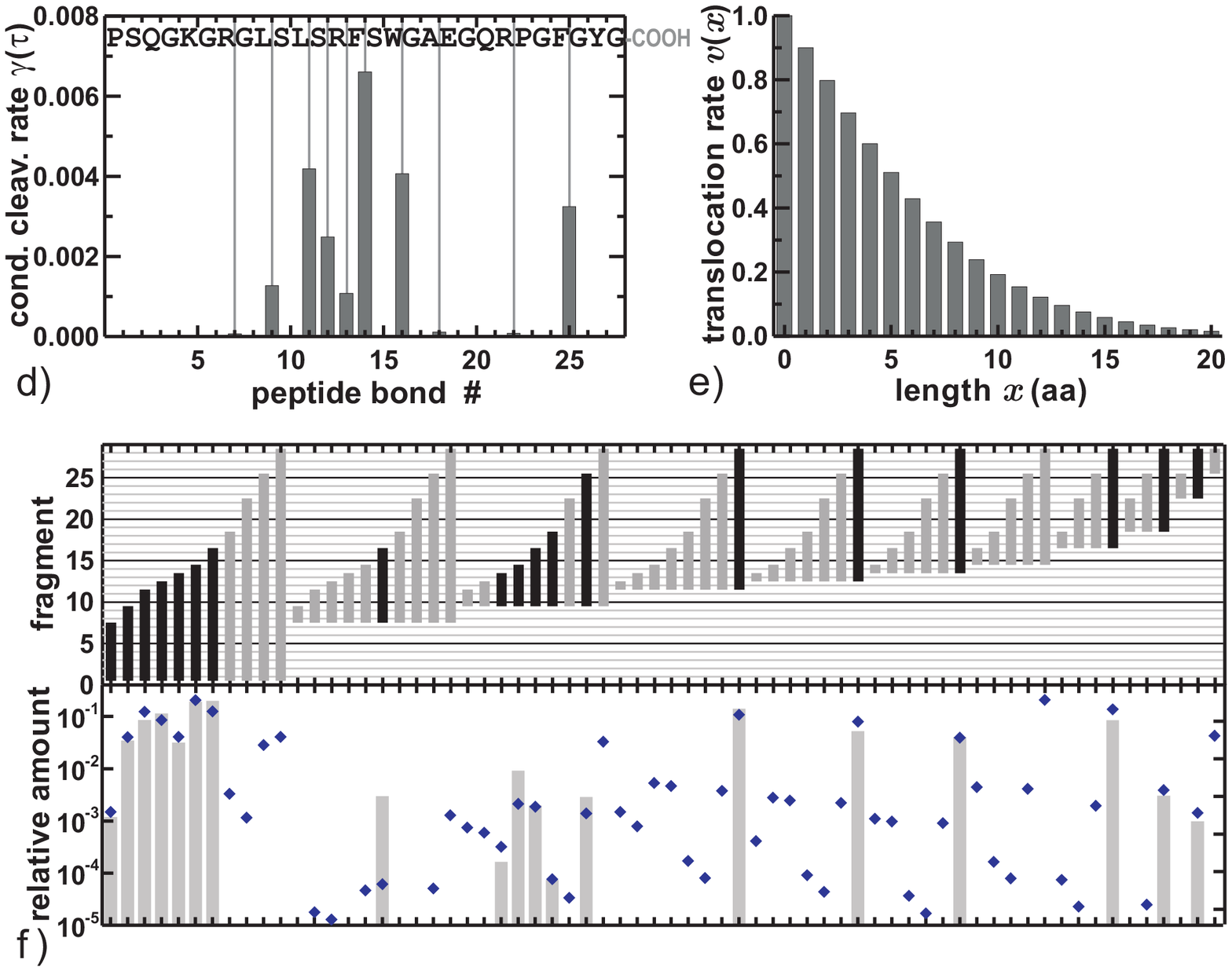}
}
  \caption{Test (a--c): Reconstruction of translocation function
$v_x$ and conditional cleavage rates $\gamma_\tau$ for the 28mer
peptide Kloe~320
\protect\cite{2008_Mishto_Zaikin_proteamalg}.
 a)~the conditional cleavage
probabilities and the aa sequence; b)~the translocation rate
function; c)~the upper plot presents a set of digestion fragments
(black bars: fragments utilized for the reconstruction, gray bars:
not utilized), and the lower plot presents the amount of the
corresponding fragment (diamonds: the reconstructed values
$Q_\mathrm{fin}$, gray bars: the values of $\widetilde{Q}$
utilized for the reconstruction).\\
Experiment (d--f): Reconstruction of $v_x$ and $\gamma_\tau$ for
the 28mer peptide Kloe~258 degraded by  20S proteasome;
$P_\Nt=54\%$.
}
  \label{fig2}
\end{figure}

Performing numerical simulation of degradation of $M$mer
polypeptide, one should start at $\tau=1$ with
$w(x|\tau=1)=\delta_{x,1}$ and iterate
Eqs.\,(\ref{eq01-1})--(\ref{eq01-3}) till the last $\tau=M-1$.
Additionally, the releasing of the last fragment from the chamber
at the ``time moment'' $\tau=M$ should be taken into account:
 $Q(M,M-x+1|M)\to Q(M,M-x+1|M)+w(x|M)$.
Hence, with $w(x|\tau)$ known for $\tau=1,2,...,M$, one can
summarize Eqs.\,(\ref{eq_aux-2Q}), (\ref{eq_aux-c1Q}),
(\ref{eq_aux-c2Q-1}) and (\ref{eq_aux-c2Q-2}) for to evaluate the
digestion pattern, {\it i.e.}\ the total amount $Q(m,n)$ of the
peptide $(n,m)$ generated during a single polypeptide processing,
\begin{eqnarray}
 &&\hspace{-4mm}
 Q(\tau_1,\tau_2)=Q(\tau_1,\tau_2|\tau_1)
 +\Theta(M\!-\!\tau_1\!-\!L)\,Q(\tau_1,\tau_2|\tau_1\!+\!L)
 \nonumber\\[5pt]
 &&\!=\delta_{\tau_1,M}w(\tau_1+L-\tau_2+1|M)
 \nonumber\\[3pt]
 &&\quad
 {}+\frac{\gamma_{\tau_1}\,w(\tau_1-\tau_2+1|\tau_1)}
 {v_{\tau_1-\tau_2+1}+\gamma_{\tau_1}+\Theta(\tau_1\!-\!\tau_2\!-\!L)\gamma_{\tau_1-L}}
 \nonumber\\[0.5pt]
 &&\quad
 {}+\frac{\delta_{\tau_1-\tau_2+1,L}\,\gamma_{\tau_1}}{v_L+\gamma_{\tau_1}}
 \!\sum_{x=L+1}^M\frac{\gamma_{\tau_1-L}\,w(x|\tau_1)}{v_x+\gamma_{\tau_1}+\gamma_{\tau_1-L}}
 \nonumber\\
 &&
 {}+\Theta(M\!-\!\tau_1\!-\!L)
 \frac{\gamma_{\tau_1}\,w(\tau_1+L-\tau_2+1|\tau_1+L)}
 {v_{\tau_1+L-\tau_2+1}+\gamma_{\tau_1+L}+\gamma_{\tau_1}},
 \label{eq03}
\end{eqnarray}
here $1\le\tau_2\le\tau_1\le M$. Since the protein can be cleaved
starting both from the C- and from the N-terminal, the final
digestion pattern is given by
\begin{eqnarray}
&&\hspace{-1cm}Q_\mathrm{fin}(\tau_1,\tau_2)=P_\Nt\,Q_\Nt(\tau_1,\tau_2)\nonumber\\
 &&+P_\Ct\,Q_\Ct(M-\tau_2+1,M-\tau_1+1)\,.\label{eq04}
\end{eqnarray}
The subscripts indicate which terminal goes first, $P_\Nt$ and
$P_\Ct=1-P_\Nt$ are the probabilities of the degradation starting
from the corresponding end.

Digestion pattern $Q_\mathrm{fin}(\tau_1,\tau_2)$ is a functional
of TRF $v_x$ and CCR $\gamma_\tau$. Utilizing MS data on the
digestion pattern, one can determine nonzero values of
$\gamma_\tau$ ({\it i.e.}\ positions of possible cleavage) and
minimize the mismatch between $Q_\mathrm{fin}(\tau_1,\tau_2)$ and
MS data $\widetilde{Q}(\tau_1,\tau_2)$ over $v_x$, the nonzero
values of $\gamma_\tau$, and $P_\Nt$ in order to {\it reconstruct}
them.
Note, $v_x$ and $\gamma_\tau$ are defined up to the constant
multiplier, which should be determined from the degradation rate
in real time, but not from the digestion pattern.

In order to verify the robustness of the reverse engineering
procedure, numerous tests have been performed. A typical test
presented in Figs.\,\ref{fig2}a--c has been performed as follows.
For given $v_x$
and $\gamma_\tau$ the digestion pattern $Q(\tau_1,\tau_2)$ has
been evaluated. The result has been perturbed by noise;
 $\widetilde{Q}(\tau_1\tau_2)=Q(\tau_1\tau_2)
 +10^{-4}R_{\tau_1,\tau_2}\sqrt{Q(\tau_1,\tau_2)}$,
where $R_{\tau_1,\tau_2}$ are independent random numbers uniformly
distributed in $[-1,1]$. The information about 1mer and 2mer
fragments and fragments which relative amount is less than
$5\cdot10^{-3}$ has been omitted as being hardly acquirable in
experiments
\cite{2001_Kohler}. Resulting $\widetilde{Q}_{\tau_1\tau_2}$ has
been used for the reconstruction of $v_x$ and $\gamma_\tau$. The
original and reconstructed data for $\gamma_\tau$
(Fig.\,\ref{fig2}a) and $v_x$ (Fig.\,\ref{fig2}b) are in a good
agreement. The reconstructed $P_\Nt=0.49$ against the original
$P_\Nt=0.50$.

%

Figs.\,\ref{fig2}d--f present the results of the reverse
engineering from the experimental ({\em in vitro}) digestion
pattern for the 28mer Kloe~258, which is the sequence 101--128\,aa
of human Myelin Basic Protein, degraded by 20S proteasomes
purified from lymphoblastoid cell lines, which express mainly the
immunoproteasome (for materials and methods
see~\cite{2008_Mishto_Zaikin_proteamalg}).
 The TRF $v_x$ appears to be
monotonically decaying; the reconstructed probability of the
degradation starting from the N-terminal $P_\Nt=54\%$, meaning the
degradations from the N- and C-ends are almost equally probable in
this case.

The suggested reconstruction method has some limitations. The
reconstruction procedure for short polypeptides is very sensitive
to measurement inaccuracy. Though the whole information on
$Q(\tau_1,\tau_2)$ is not needed, the number of nonzero values of
$Q(\tau_1,\tau_2)$ utilized for a reliable (tolerant to noise)
reconstruction
 should considerably exceed the number of reconstructed
 parameters.
For Kloe~258 the number of trustworthy and utilized values of
$\widetilde{Q}(\tau_1,\tau_2)$ is $19$ (see Fig.\,\ref{fig2}f), it
is a bit greater than the number of unknown parameters which is
14. Hence, more accurate and comprehensive MS data on the
digestion pattern are needed. Additionally, for short polypeptides
the finishing stage of the degradation is relatively important,
because on this stage the translocation rate is affected by the
edge effects (the backward end of the polypeptide gets inside the
proteasome chamber) and is not the same as for the remainder of
the polypeptide.

%

Fast and effective design of new intelligent drugs against immune
and autoimmune deceases~\cite{2001_Kloetzel_nature} is impossible
without development of the virtual immune system, by which these
drugs can be tested {\it in silico}. One of the most important
steps on this road is the prediction of presentation profile, {\it
i.e.} number of epitopes,
from  transcription to presentation on MHC class I complex and potentially  recognized by CD8+ T-cells. To do this, one should be
able to predict reliably the proteasomal digestion pattern (DP)
that is determined by sequence-specific cleavage preferences
(SSCP), that is quantified by CCR in our approach, and proteasomal
TRF. Some attempts to fulfill these predictions have been made
based on finding the correlations between SSCP and final DP. These
algorithms are available on the
Internet~\cite{2000_Kuttler_Paproc,2005_Tenzer} however they are
not always reliable because of ignoring the dependence of DP on TRF. The
significance of the mathematical method presented here is the
possibility to find dependencies between SSCP, TRF and DP. Using
this method and applying it to various experimental data one would
be able to construct algorithms for a reliable prediction of
proteasomal DP.

In summary, we have proposed a model of the degradation of
proteins by the proteasome which allows one to {\em reconstruct}
the proteasomal transport function and cleavage strengthes. The
model is applicable to a broad variety of hypothetically possible
translocation
mechanisms~\cite{2005_Zaikin_proteasome,2001_Reimann}. We have
tested the model for relatively short (25--50mers) synthetic
polypeptides as the most common case for {\it in vitro}
experiments.
Earlier, in~\cite{2006_Zaikin_Goldobin}, we have  described how
peculiarities of the translocation function may lead to the
multimodality of the fragment length distribution even for
$\gamma(\tau)\equiv const$. Here we have shown that the amount of
each digestion fragment is not only determined by the cleavage map
of the substrate but is also crucially affected by nonuniformity
of the translocation rate. The proposed methodology can be used in
extensive analysis of already available MS data for the 20S
proteasomes and its associations with different regulatory
complexes and under different experimental conditions. The results
of this analysis, specifically, the shape of the translocation
rate function and its variations for diverse proteasome species
under different conditions, can give insight into the mystery of
the protein translocation mechanism inside the proteasome. Such an
analysis can elucidate also the unanswered question whether there
is some preference for starting the degradation with the N- or
C-terminal of the protein, and how this preference is
quantitatively affected by regulatory complexes.


We thank S.\ Witt for fruitful discussions and the VW-Stiftung, PROTEOMAGE  (FP6), the BRHE program, and
``Perm Hydrodynamics'' for financial support.


\end{document}